\documentclass[]{article}
\usepackage{authblk}
\usepackage{graphicx}
\usepackage{float}
\usepackage{listings}
\usepackage{subcaption}
\usepackage{amsmath}
\usepackage{amssymb}
\usepackage{soul}

\usepackage{color}
\definecolor{red}{rgb}{1,0,0}
\definecolor{blue}{rgb}{0,0,1}

\begin{document}

\title{
{Bose condensation and Bogoliubov excitation in resonator-embedded superconducting qubit network}  
}

\author[1]{Patrick Navez}
\author[2]{Valentina Di Meo}
\author[3]{Berardo Ruggiero}
\author[4]{Claudio Gatti}
\author[5]{Fabio Chiarello}
\author[4]{Alessandro D'Elia}
\author[4]{Alessio Rettaroli}
\author[6]{Emanuele Enrico}
\author[6]{Luca Fasolo}
\author[7]{Mikhail Fistul}
\author[7]{Ilya Eremin}
\author[8]{Alexandre Zagoskin}
\author[3]{Paolo Vanacore}
\author[9]{Paolo Silvestrini}
\author[2]{Mikhail Lisitskiy}

\affil[1]{Université de Montpellier}
\affil[2]{Institute for Superconductors, Innovative materials, and devices - CNR SPIN}
\affil[3]{Institute of Applied Sciences and Intelligent Systems - CNR ISASI}
\affil[4]{National Laboratory of Frascati LNF, National Institute of Nuclear Physics - INFN}
\affil[5]{Institute for Photonics and Nanotechnologies - CNR IFN}
\affil[6]{National Institute of Metrological Research, INRiM}
\affil[7]{ Theoretische Physik III, Ruhr-Universit\"at Bochum, Bochum 44801, Germany}
\affil[8]{Loughborough University}
\affil[9]{Department of Mathematics and Physics, University of Campania "Luigi Vanvitelli"}

\maketitle

\begin{abstract}
Superconducting qubit networks (SQNs) embedded in a low-dissipative resonator is a promising device allowing one not only to establish the collective quantum dynamics on a macroscopic scale but also to greatly enhance the sensitivity of detectors of microwave photons. A quantum ac Stark effect provided by coupling between an SQN and microwave photons of a resonator, leads to a strong nonlinear interaction between photons.  Here, we present a two-tone spectroscopy experiment in which a set of 10 superconducting flux qubits is coupled to the input $R$- resonator and the output $T$- transmission line. 
An external  microwave pump field  close to the resonance frequency populates macroscopically  the resonator mode  as  a Bose-Einstein condensate, while  a second probe beam scans the resonances referred also as Bogoliubov-like excitations.
The corresponding excitation frequency  measured from the transmission coefficient, $|S_{21}(f)|$ displays  an abrupt change of the resonant dip position  once the power of the  pump field overcomes a critical value $P_{cr}$. This sharp shift occurs in a narrow region of pump frequencies, and can be tuned by an applied magnetic field.  It is a signature of bistability of the photon number inside the resonator, in agreement with theory.


\end{abstract}

\section*{Introduction}
Josephson junctions  circuits whose dynamics are governed by a large set of nonlinear equations of motion, have been used for many years as a unique playground allowing one to study plenty of fascinating physical phenomena such as magnetic Josephson vortices (fluxons) \cite{UstCirMal93,Ust98},
discrete breathers in Josephson junction ladders \cite{Trias00,Binder00},
superconductor-insulator quantum phase transitions \cite{Haviland00,Mooij15}, to realize microwave induced dynamic metastable states  and abrupt multiple switching between them \cite{Siddiqi05,Lazaridis13,Jung14}. The latter  multistability effect already observed in single Josephson junctions \cite{Wallraff03}, trapped magnetic fluxons \cite{Fistul00,Fistul03}, and in arrays of RF-SQUIDs embedded in a low-dissipative resonator \cite{Jung14,Kiselev19}, exploits the intrinsic tunable nonlinearity of Josephson junctions circuits and a reliable inductive coupling of such circuits to an electromagnetic environment.

The explosive growth of quantum technologies since the turn of the century led to the development and detailed investigation of a variety of superconducting qubit networks (SQNs) \cite{Krantz2019,Kjaergaard20}. Their dynamics is governed by the quantum-mechanical laws on a macroscopic scale, and it was used to realize different quantum information devices, e.g., quantum processors, quantum annealers, sensitive photon detectors \cite{Acin2018,Wu21,Boxio14,Yark22,Schuster07,Wang21,Enrico23}.
Even a weak coupling of SQNs to a {high-quality resonator
a stable interaction between well separated qubits of SQN} produces a non-negligible qubit-qubit coupling via resonator modes \cite{Bias21,Brehm21,Fistul22} leading to the appearance of collective quantum states of SQNs. Such collective quantum states have been observed in an array of flux/transmon qubits \cite{Macha14,Shulga17}. Due to a seminal ac Stark effect the qubit frequencies vary with the number of microwave photons of a resonator, and as a back influence a weak Kerr nonlinearity and interaction between microwave photons can be established \cite{Puri17,Maleeva18}. The ac Stark effect was used to resolve different Fock states of microwave photon field \cite{Schuster07}.
Moreover, in the Refs. \cite{Fistul22,Knap10,Mukhin13} it was predicted that the coupling of a large set of qubits with a low-dissipative high-quality resonator can lead to  collective non-classical states of microwave photon field.

In this work, we study the non linearity of a resonator photon mode pumped by a powerful 
input microwave radiation. The macroscopic population of this mode behaves like a Bose-Einstein condensate when it interacts with an SQN coupled to two low-dissipative resonators with an architecture similar to \cite{Fistul22,Enrico23,qn}.  We report on the observation of the Bogoliubov-like collective excitations of the second tone probe of a close frequency which displays bistability.  We derive a simple tractable time-dependant non-linear Schrodinger-like equation  describing the spectroscopy caused by SQN embedded in a resonator and which  reproduces  quite remarkably the experimental data. In previous work \cite{Bishop, Kurko,Blais,Sett}, bistability has been already predicted and observed but only for a unique qubit.


\begin{figure} [H]
    \centering
    \includegraphics[width=0.85\linewidth]{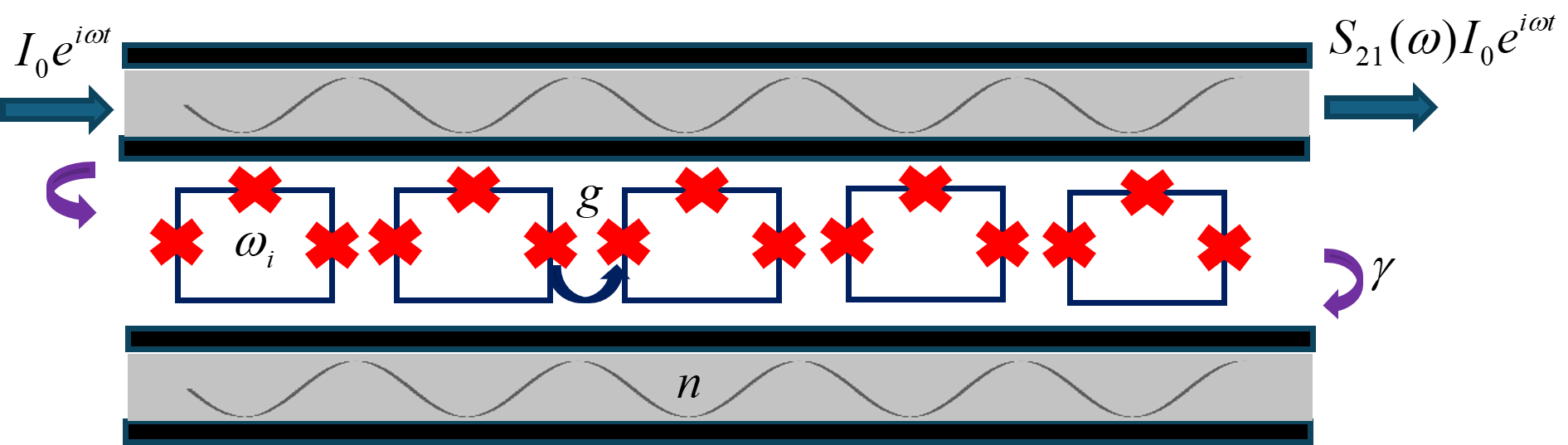}
    \caption{Schematic of a proposed device: an SQN composed of $N$ flux qubits coupled to a low-dissipative resonator (bottom) and a transmission line (top).  }
    \label{fig:1p}
\end{figure}

\section*{Results}

\subsection*{Physical device and experimental setup }
The implementation of the device of Fig. \ref{fig:1p}) involves two resonators ($T$-resonator and $R$-resonator) coupled through an SQN. The latter consists of 10 capacitively-shunted superconducting flux qubits, each one containing a superconducting loop interrupted by 3 Josephson junctions \cite{Mooji1998, Orlando1999}. The SQN was coupled inductively to the $T$-resonator and capacitively to the $R$-resonator. The  \textcolor{blue}{had a} coupled resonant frequency $f_r \simeq 7.7 GHz$ and the coupling quality factor $Q_c \simeq 10^5$.

\begin{figure}[t!]
    \centering
    \begin{subfigure}{0.75\textwidth}
        \includegraphics[width=\textwidth]{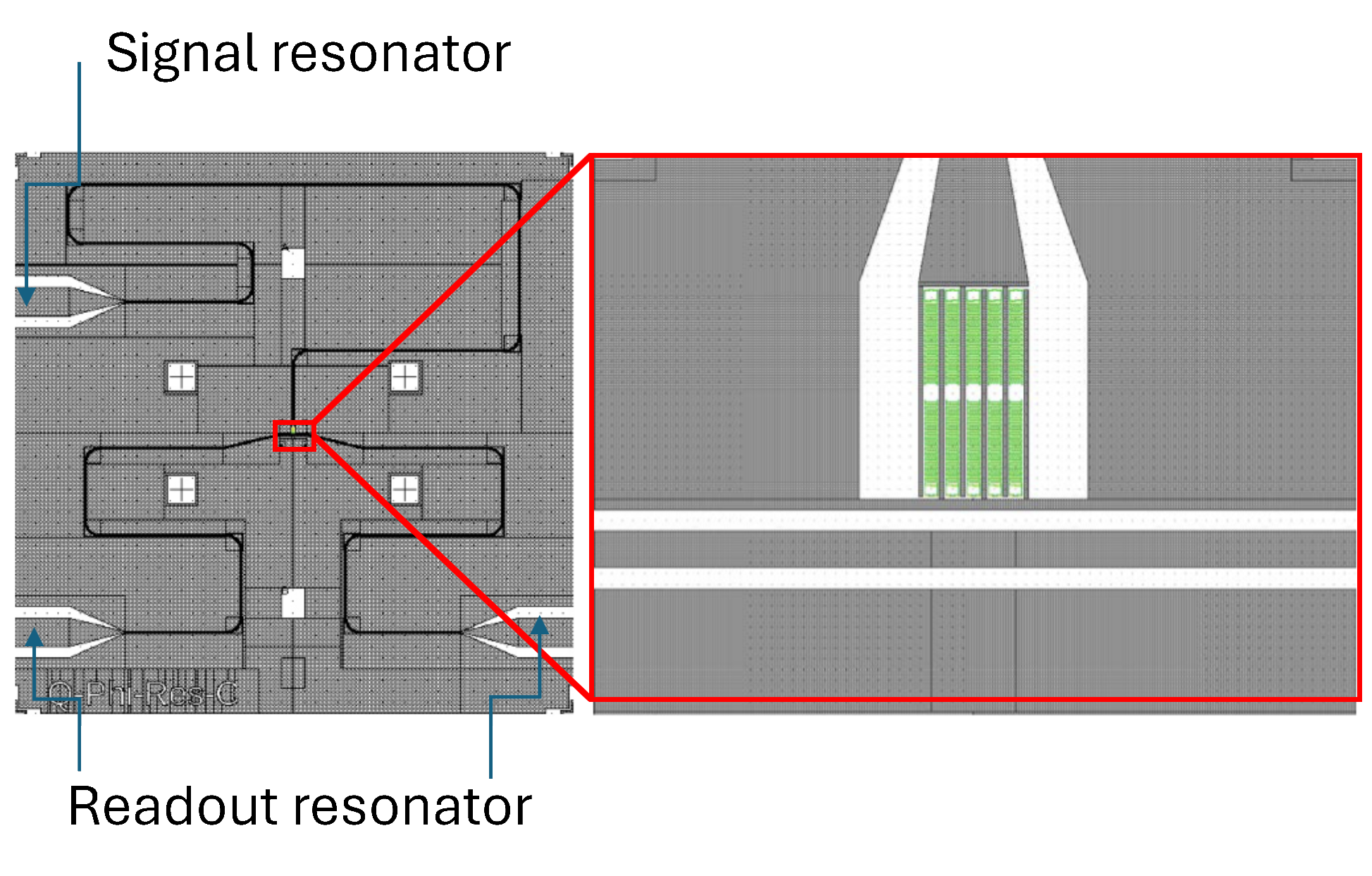}
        \caption{}
        \label{fig:lay_a}
    \end{subfigure}
      \vspace{0.5cm}
    \begin{subfigure}{0.5\textwidth}
        \includegraphics[width=\textwidth]{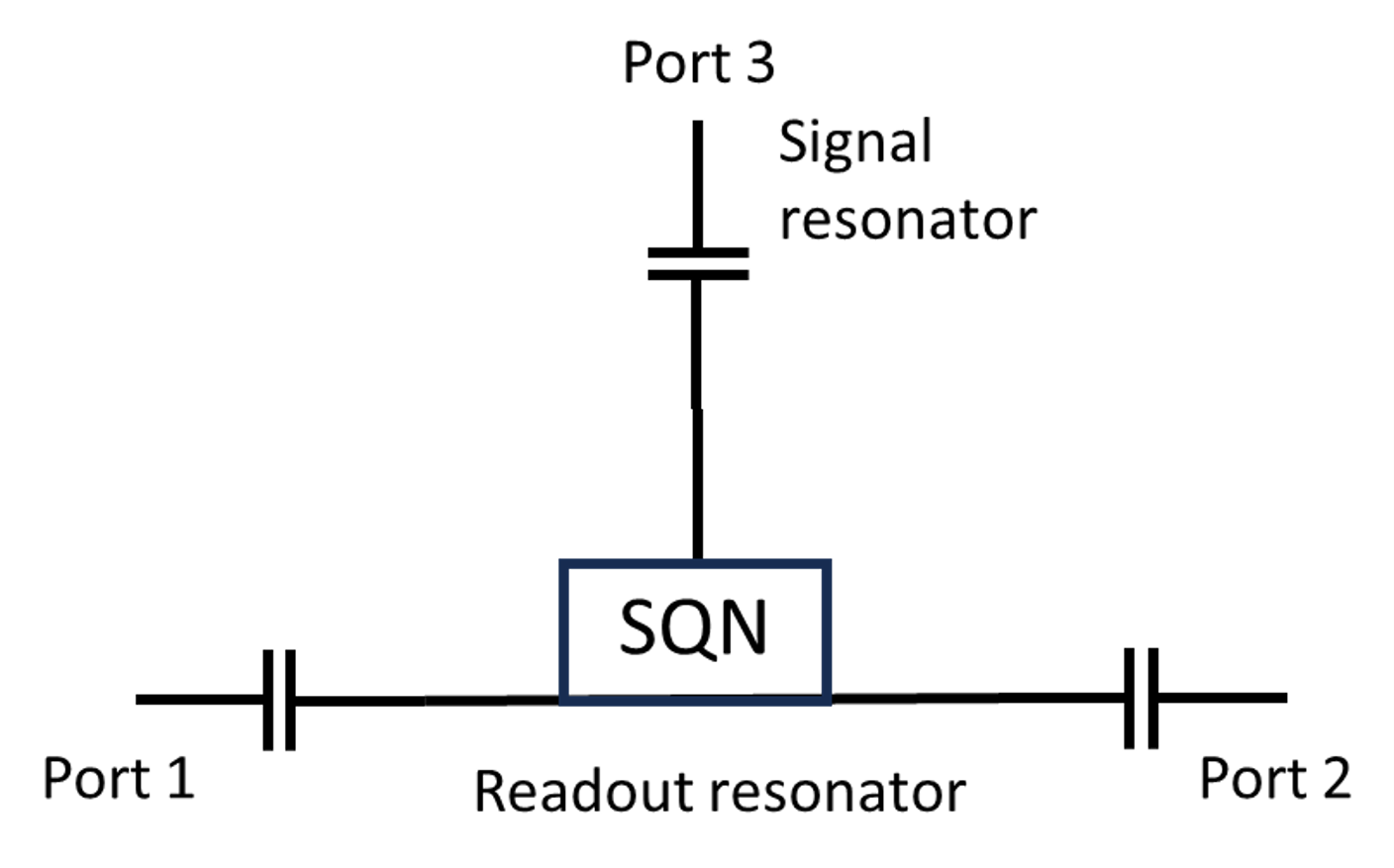}
        \caption{}
        \label{fig:scheme_c}
    \end{subfigure}
    \caption{\textbf{Device layout and experimental setup.} 
    (a) (Left) Layout of the device. (Right) The optical micrograph of 10 capacitively-shunted flux qubits integrated with $R$- resonator and $T$-transmission line. Port numbers are indicated.
    (b) Schematic of the experimental setup highlighting the microwave paths and components. }
    \label{fig:2p}
\end{figure}

The $T$- and $R$- resonators are delimited by two 5~fF capacitors, and a 5~fF capacitor on a one side and 5 flux qubits on the other side, accordingly. Such configuration allows one to strongly suppress 
unwanted parasitic coupling between the two resonators in absence of a microwave photon field, and ensure their mutual electrodynamic isolation in absence of the SQN. The device layout and the optical image of the SQN coupled to the $T$- and $R$- resonator are shown in Fig.~\ref{fig:lay_a}. The standard characterization of the fabricated devices and experiments were carried out at 15 mK temperature in a Leiden Cryogenics CF-CS110-1000 dilution refrigerator installed at the Laboratori Nazionali di Frascati (LNF, Italy).
The details of the fabrication process of the resonators and the SQN is reported in the Materials and Methods Sec. A.

We carried out two-tone spectroscopy experiment in which the transmission coefficient $S_{21}(f_{p})$ between ports 2 and 1 of the the $T$-resonator in the presence of microwave photons injected at port 3 of the $R$-resonator, was measured. Here, $f_{p}$ and $f_{pump}$ are the frequencies of probe (the first-tone) and pump (the second-tone) signals, accordingly. The frequency $f_{pump}$ was varied in a narrow region around the resonant frequency of the $R$-resonator. For spectroscopic measurements an Agilent E5071C Vector Network Analyser (VNA) was used. The majority of experiments were carried out in the absence of external magnetic field. The power of the second-tone signal was varied over the range –110 dBm to –60 dBm. 
Schematic of the overall experimental setup is shown in Fig.~\ref{fig:2p}b (details are presented in Material and Methods Sec. B).

\subsection*{Two-tone spectroscopy: Measurements of the Transmission Coefficient}

\begin{figure}[]
 \centering
    \includegraphics[width=\linewidth]{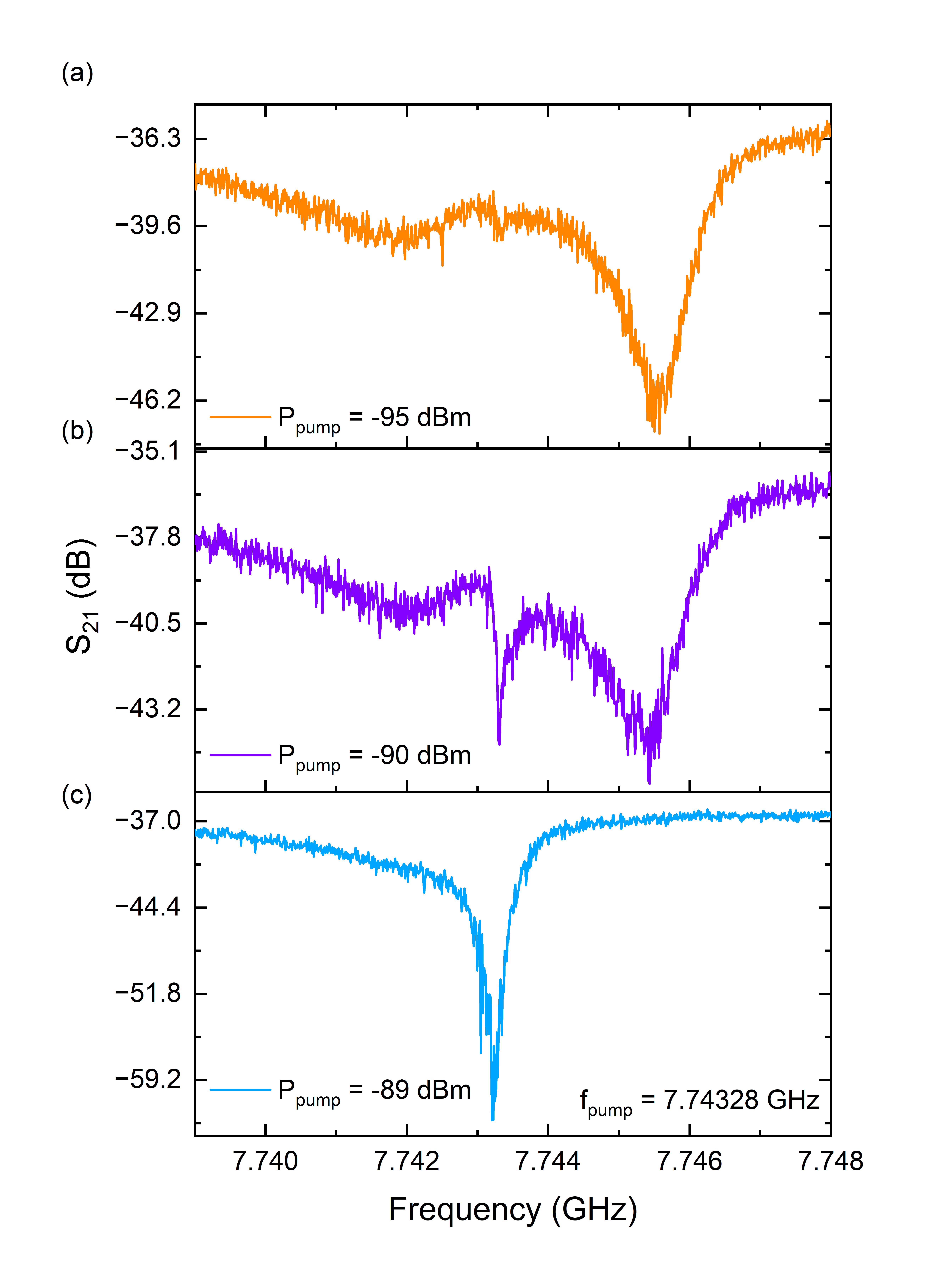}
    \caption{The dependence of the transmission coefficient $|S_{21}|$ on the probe frequency $f_{p}$ for three different values of pump signal power: $P_{pump}<P_{cr}$(a); $P_{pump} \simeq P_{cr}$ (b); $P_{pump} \geq P_{cr}$ (c). The frequency of the pump signal was 7.74328~GHz. The externally applied magnetic field is zero. Note the weaker mirror idler resonance at about 7.742 GHz in graph (a) and (b) corresponding to the second branch of the Bogoliubov excitation.}
 
 \label{fig:3}
\end{figure}

 The typical measured dependencies of the $|S_{21}|$ on the frequency of a probe signal, $f_{p}$, demonstrating resonant dips, are shown in  Fig.~\ref{fig:3} for three slightly different values of pump signal power. The observed resonant dips are very narrow ones indicating a high coupling quality factor $Q_c$ of the system. 
 
 Our main experimental observation is a clear \textit{shift of the resonant dip position} toward lower frequencies as the power of the second-tone signal increases. The shift becomes particularly pronounced as the pump signal power $P_{pump}$ exceeds a threshold power $P_{cr}$. We observe also a substantial narrowness of the resonant dip as $P_{pump}$ was slightly above the $ \simeq P_{cr}$ (see Fig.~\ref{fig:3}b).
 
\begin{figure}
 \centering
 \includegraphics[width=\linewidth]{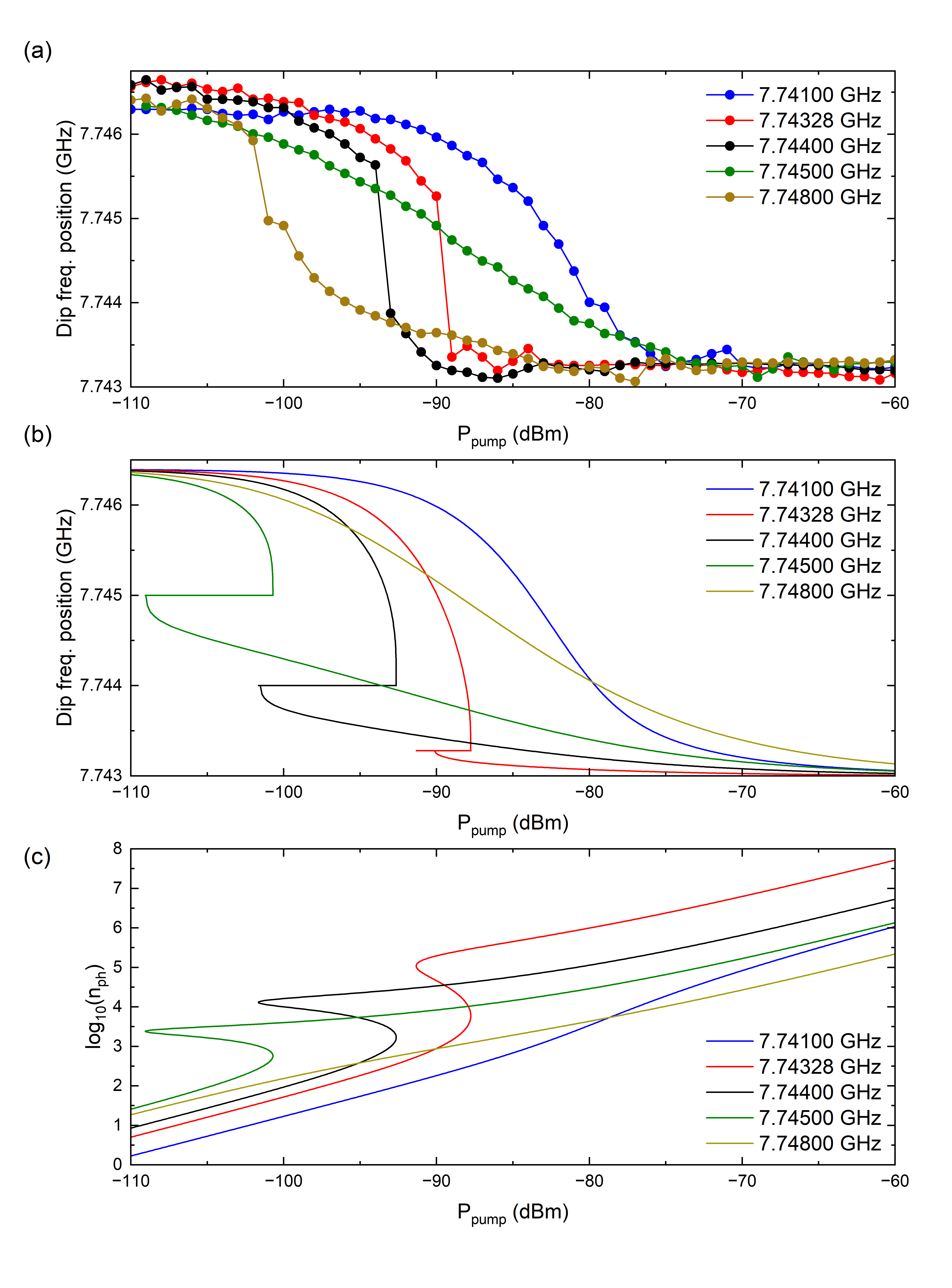}
 \caption{(a) The experimental  dependence of the dip frequency position on the pump power $P_{pump}$ for different frequencies of the pump signal.(b) Calculated dependencies of the dip frequency position on the pump power $P_{pump}$ for different frequencies of the pump signal. (c) Deduced photon number inside the cavity. To obtain the best fit to the experimental curves we choose the following parameters as: $\omega_c=2\pi \times  7.7430 GHz$, $\gamma_c=2\pi \times 3.2 MHz$, $\Delta=2\pi \times 1.78\,GHz$ and the coupling $g=2\pi \times 25\, MHz$. The  dashed line corresponds to the real line as seen in experiment. Within the bistability region,  the photon number in the resonator is in  the  lower branch  in the green and black curves while  it is in  the higher branch in the red curve. 
}
 \label{fig:4}
\end{figure}

  The dependence of the dip frequency position on $P_{pump}$ is presented in Fig.~\ref{fig:4}a for different values of pump signal frequencies, $f_{pump}$. The position of the resonant dip strongly depends on pump signal frequency in a narrow window of pump frequencies,  $7. 743 GHz < f_{pump}< 7. 746$, only. Moreover, the magnitude of a sharp drop decreases and the pump power threshold $P_{cr}$ shifts downwards as the pump frequency was slightly increased. 

\begin{figure}[H]
\includegraphics[width=0.85\linewidth]{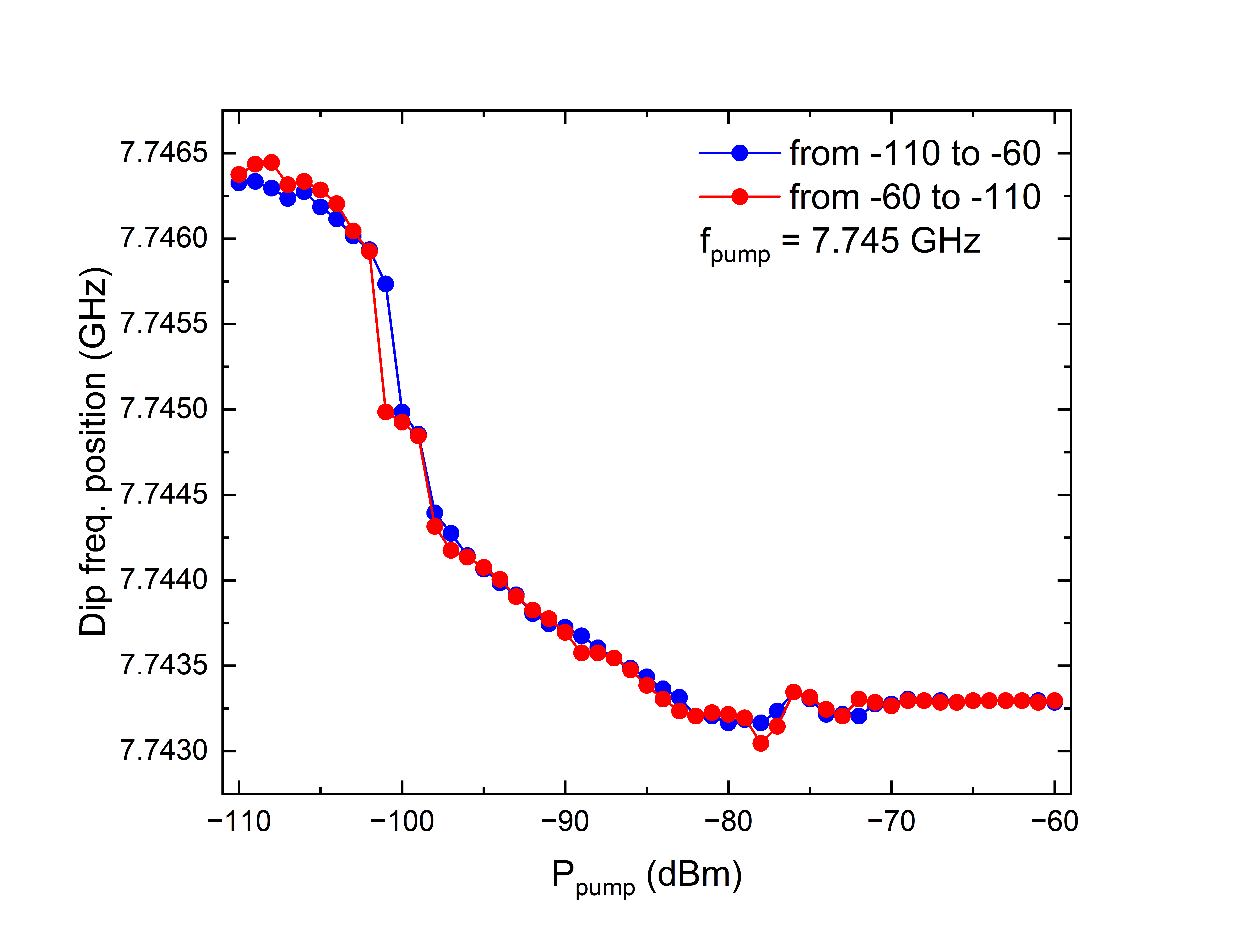}
\caption{The dependence of the dip frequency position on the pump signal power $P_{pump}$ for different directions of power sweep. 
The frequency of a pump signal was 7.745~GHz
}
\label{fig:5p}
\end{figure}

In this range of pump frequencies several observations indicate the presence of a \textit{bistable regime} in the microwave photon field. First, the frequency position of the resonant dip does not evolve continuously but exhibits an abrupt change 
(see Fig.~\ref{fig:4}) as the pump power sweeps from –110 dBm to –60 dBm.
Second, the dependencies of the transmission coefficient on the probe frequency shown in  Fig.~\ref{fig:3} reveal the emergence of a secondary resonant feature, which develops as the pump power $P_{pump}$ approaches the threshold $P_{cr}$ (Fig.~\ref{fig:3}b). Once $P_{pump} \geq P_{cr}$ this secondary resonant dip dominates the transmission spectrum (Fig.~\ref{fig:3}c), confirming that the system has entered a new metastable configuration. Another evidence for the bistability region is a small hysteresis shown in Fig.~\ref{fig:5p}. For $P_{pump} \simeq P_{cr}$ the resonant dip position depends also on the the direction of the pump signal power sweep. When the second-tone power increases from –110 dBm to –60 dBm and subsequently decreases back to –110 dBm, the dip position trajectory forms a hysteretic loop.

\begin{figure}[H]
    \centering
    \includegraphics[width=0.85\linewidth]{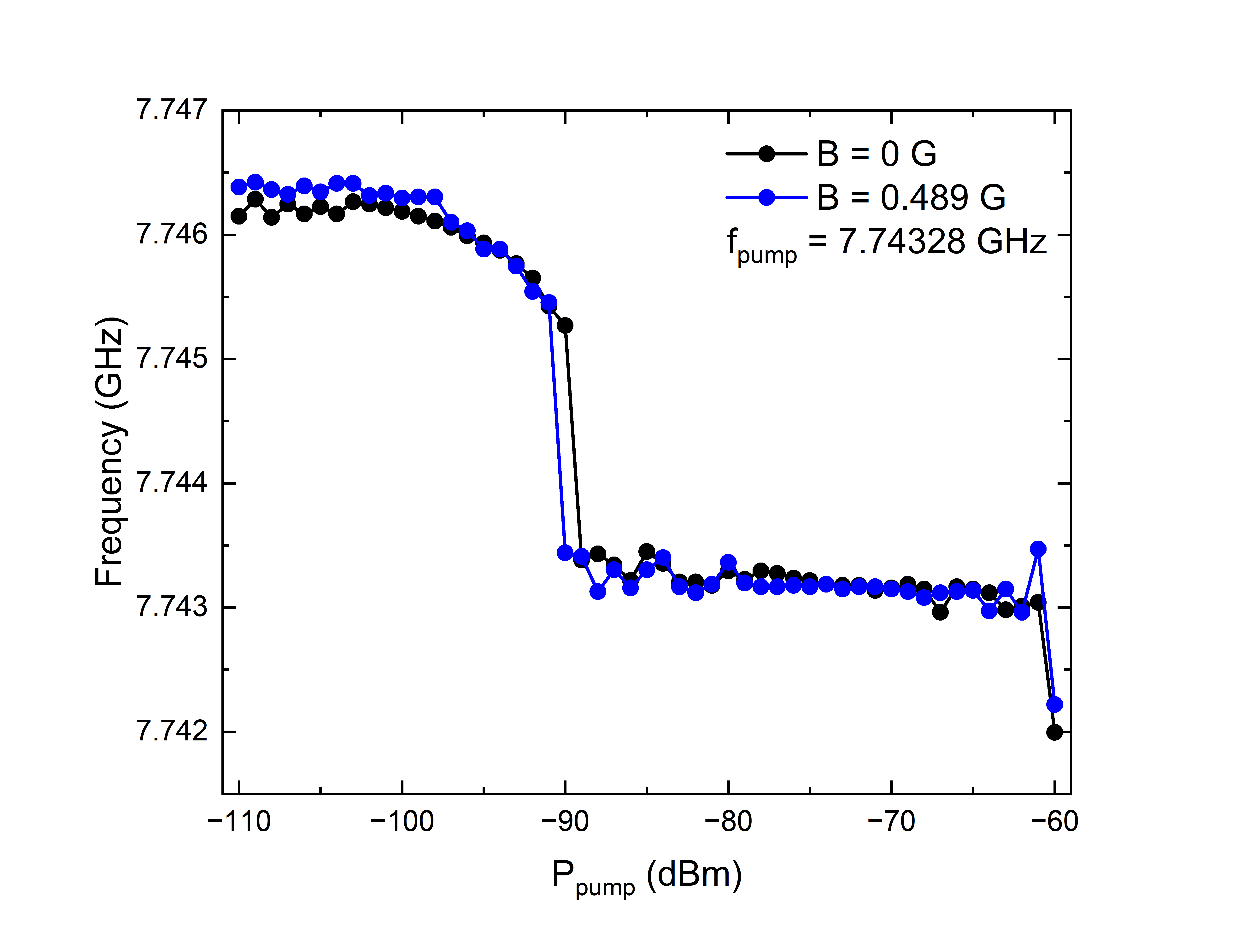}
    \caption{The dependence of the dip frequency position on the pump power in the presence of a static magnetic field, $B=0.489 G$,
    The pump frequency was $f_{pump}=7.74328$.
    }
    \label{fig:6p}
\end{figure}

We also performed the two-tone spectroscopy measurements under a static magnetic field of $B = 0.489~ G$. 
The results presented in Fig.~\ref{fig:6p}, indicate a shift of the pump power threshold $P_{cr}$ to a lower values as a static magnetic field was applied.



\subsection*{Theory: Collective Bogoliubov modes of Microwave Photons }

{\bf General analysis:} 
We treat the structure as a single resonator with the resonance frequency $\omega_c$ produced by the hybridisation of resonance modes of the R- and T- resonators.
The two-tone spectroscopy measurements of an SQN composed of $N$ qubits with the frequencies $\omega_j$ and coupled to high-quality resonators is quantitatively described by a time-dependent nonlinear Gross-Pitaevskij-like equation for the photon field inside the $R$+$T$-resonator, $\psi(t)$, written in the system rotating with the pump frequency $\omega_0$ and normalized such that the average photon number is $n_{ph}= |\psi(t)|^2$ (see Materials and Methods, Part C). In presence of an external microwave input represented by the canonical variable $x_{in}(t)$, we find:
\begin{eqnarray}\label{DynamicsEq}
 \left[i\frac{d}{dt} +\omega_0-\omega_c -\sum_{j=1}^N \frac{g_j^2/\Delta_j}{\sqrt{1+4g^2_j|\psi(t)|^2/\Delta^2_j}}+i\frac{\gamma_c}{2}\right]\psi(t)= 
 \sqrt{\frac{\omega_c}{2|Q_c|}}\frac{x_{in}(t) }{i}
\end{eqnarray}
where $\omega_0=2\pi f_{pump}$ is  the pump  frequency; $\Delta_j=\omega_0-\omega_j$ and $g_j$ is the coupling of the $j$-th qubit to the $R+T$-resonator. We also introduce the $R+T$-resonator dissipation, $\gamma_c=\omega_0/Q_c$.
The interaction between photons is produced by the Stark shift of qubits frequencies induced by the coupling between an SQN and microwave photons. It displays a non-linearity quite different from the Gross-Pitaevskii equations since it  disappears in the limit of strong field restraining its effect in a delimited range of power.

In the presence of pump microwave input only $x_{in}(t)=x_0$ with power  $P_{pump}=\hbar \omega_c|x_0|^2$, we obtain a constant complex solution providing the total photon number inside the resonator:
\begin{eqnarray}\label{nph}
n_{ph}
=\frac{2 P_{pump}/( \hbar |Q_c|)}
{\gamma_c^2+4\omega_s^2},
\end{eqnarray}
where we define  the effective detuning:
\begin{eqnarray}\label{os}
\omega_s=\omega_{0}-\omega_r\quad \quad \omega_r=\omega_c +\sum_{j=1}^N \frac{g_j^2/\Delta_j}{\sqrt{1+4g^2_jn_{ph}/\Delta^2_j}}.
\end{eqnarray}
Notice here, that the condition $\omega_s=0$ determines the frequency position of the narrow resonant drop in a \textit{single tone } spectroscopy measurements, i.e., the dependence $|S_{33}(\omega_0)|$ determined from Eq.\eqref{eq:luca}. 

In two-tone spectroscopy measurements where both the pump and probe microwave fields are present, one can take a low amplitude perturbation in the following form: $x_{in}(t)=x_0+\delta x(t) =x_0+x_p\exp[-i(\omega_{p}-\omega_0)t]$ on the constant pump signal, $x_0$, and the perturbative probe signal $x_p$ leading to the photon number perturbation $\delta \psi(t)=e^{-i(\omega_{p}-\omega_0) t} u +e^{i(\omega_{p}-\omega_0) t} v$ {which, in addition to the probe frequency $\omega_p$, contains another idler mode  at frequency $2\omega_0-\omega_p$}. 
In this case the solution of the Eq. (\ref{DynamicsEq}) determines the eigenfrequency of the Bogoliubov-like collective excitations induced by the perturbation: 
\begin{eqnarray}\label{sh}
\omega_{p}=\omega_0 -\omega_s {\rm Re}\sqrt{1+  \sum_{j=1}^N \frac{4 g_j^4 n_{ph}/\omega_s}{(\Delta^2_j+4g^2_j n_{ph})^{3/2}}}.
\end{eqnarray}
These collective excitations correspond to the Bogoliubov modes occurring in a system of interacting photons collected in the $R$-resonator. The value $\omega_{p}$ determines the dip frequency position of the resonant drop in the $|S_{21}|$  measurements. The second branch excitation proportional to $v$ is the idler mode shown in Fig.3 justifying the Bogoliubov approach. 
The Eqs.(\ref{os}), (\ref{nph}) and (\ref{sh}) form a self consistent set to quantitatively describe the coherent collective states in the photon field trapped in the resonator. 

{\bf Bistability:}
The non linear dependancy of the power on its freque
ncy in (\ref{nph}) allows the possibility of many solutions and therefore bistability for the relation between the power and the frequency given a particular value for the Rabi frequency. However in the present case, the parameters values have not permitted to reach this regime.
Bistability is triggered once both conditions $dP_{pump}/dn_{ph}= d^2P_{pump}/dn_{ph}^2=0$ are fulfilled in Eq.(\ref{nph}).
In the case of equal detuning and coupling constant for the qubits,
we can determine
an upperbound for the cavity leakage below which bistability occurs so that two stable values of average photon number are possible given a power value.
Defining the dimensionless constant $s=1+4g^2 n_{ph}/\Delta^2$, we find using both conditions the following general bounds:
\begin{eqnarray}\label{bi}
\frac{\gamma_c}{|\omega_{0c}|} < \frac{\sqrt{3}}{2}\frac{s^2-1}{s}
\quad \quad
\left|\frac{\delta \omega_c}{\omega_{0c}}\right|> \frac{3s+ s^3}{4}
\end{eqnarray}

\begin{figure}
 \centering
 \includegraphics[width=9cm]{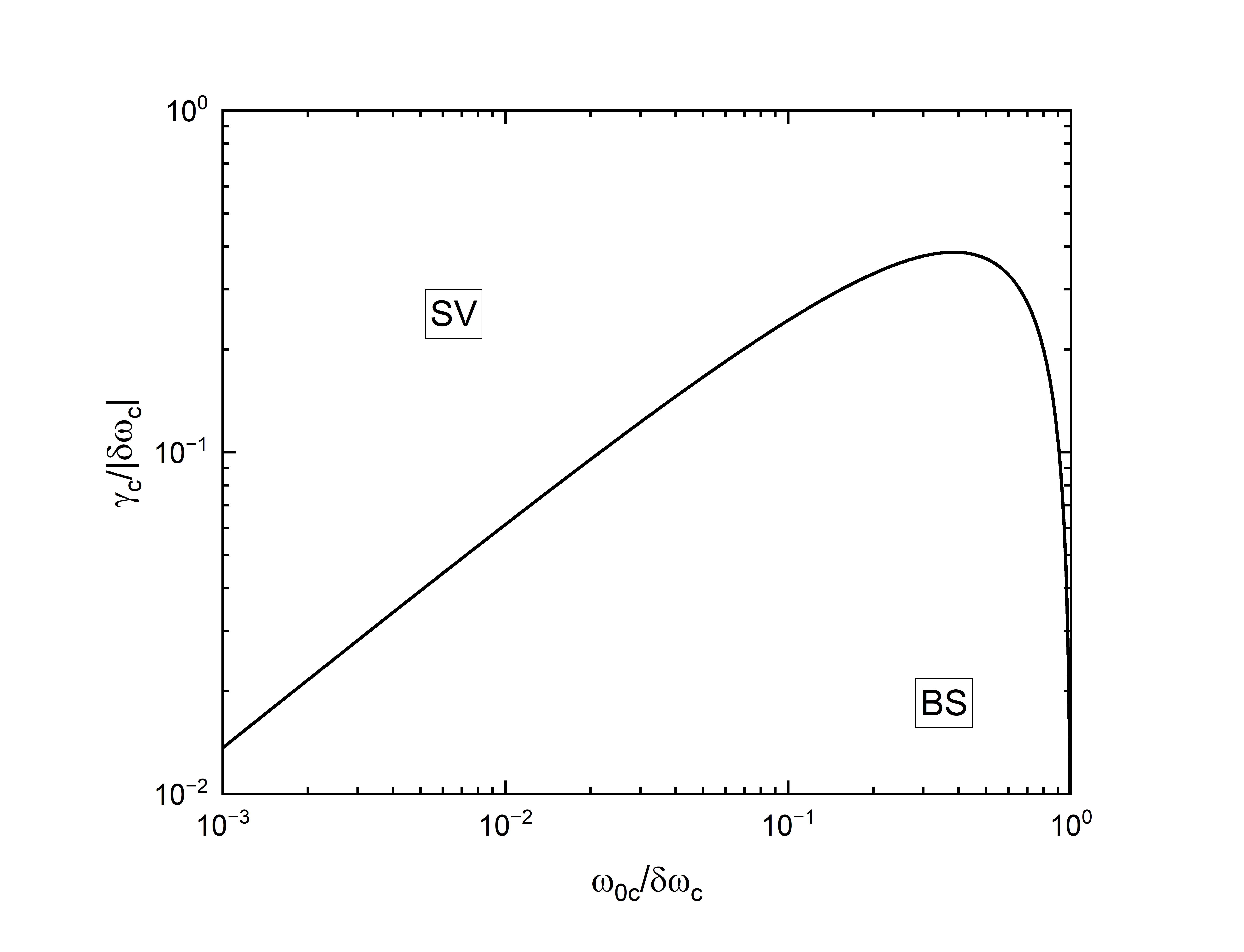}
 \caption{Phase diagram delimiting the single value (SV) region from the bistability region (BS) for any cavity shift, cavity detuning or cavity leakage ($\omega_{0c}=\omega_0-\omega_c$).}
 \label{fig:bi}
\end{figure}
The elimination of  the variable $s$ in the expressions Eqs.(\ref{bi}) leads to  a separation line beyond which the transition occurs as shown in  Fig.\ref{fig:bi}. We see that cavity detuning is essential for for realising bistability but should not be above the cavity shift $\delta \omega_c$.
Maximal  value for the cavity leakage is obtained for
$\gamma_c =|\omega_{0c}|=2|\delta \omega_c|/(3\sqrt{3})$.
The influence of the cavity leakage rate is very strong for large detuning.

\section*{Discussion}
The analysis of the Eqs.(\ref{os}), (\ref{nph}) and (\ref{sh}) demonstrates that in the regime of a low pump power $P_{pump} \simeq 0$, the frequency shift is $\delta \omega_c=\omega_r-\omega_c =\sum_{j=1}^N
 \frac{g^2_j}{\Delta_j}$, and in the opposite regime of a large pump power, i.e., as $n_{ph} \gg 1$, $\omega_{p}=\omega_c$. 
 The numerically calculated dependencies of the dip frequency position on the pump signal power, $P_{pump}$, are presented in Fig. \ref{fig:4}. Thus, one can expect the bistable regions as the pump frequency falls in the narrow window, $\omega_{c}<\omega_{p}<\omega_c +\delta \omega_c$.
 Assuming equal coupling $g_j=g$ and equal qubit detuning $\Delta_j=\Delta$, quantitative fit 
 of the six  
 experimental curves in  Fig.(\ref{fig:4}) with Eq.(\ref{sh}) and the frequency shift  $Ng^2/\Delta = 2\pi \times  3.4 MHz$ allows to extract the two remaining  parameters $g, \Delta$ using $N=10$. 
 Note that the deduced quality factor $Q_l \simeq 25000$
 corresponds rather to the coupling quality factor $Q_c$ extracted from the fit of $S_{33}$. The difference with $Q_l$ is explained  by an additional fluorescence from a strong pump causing an additional leakage (which will be the topic of a subsequent work).  
As can be seen from the comparison in Fig. \ref{fig:6p}, we obtain a good quantitative agreement between  theoretical calculations and the experimental data. We note a plateau region where
$\omega_p=\omega_0$ and the square root in Eq.(\ref{sh}) becomes imaginary.

Assuming that the magnetic field $B$ acts only on the qubit frequencies and not on the coupling, we estimate from Fig.(\ref{fig:6p}) at low power the shift variation:  $d\omega_r/dB\simeq 0.2MHz/0.489 G$. 
From Eq.(\ref{os}), we deduce the qubit frequency shift per Gauss:
$d \omega_j/ dB \simeq (\Delta/\delta\omega_c) (d\omega_r/dB) \sim 200MHz/G$ in agreement with 
expectations.
The visible dependence of critical pump power on magnetic field remains difficult to interpret as we did not observe significant change of the bistability region by tuning the qubit frequency.



In conclusion, by making use of the two-tone spectroscopy we present experimental evidences of collective states of interacting microwave photons.  The fingerprint of such collective states is a sharp shift of the frequency position of the resonant drop in $|S_{21}|$ dependence once the power of a second-tone pump signal overcomes the critical value. We identified the narrow region of pump frequencies where this phenomena can be realized. The appearance of the collective states of microwave photons was explained  in the framework of the theoretical model based on a non-linear multiphoton interaction between the pump microwave signal and qubits of the SQN with good agreement between theory and the experimental results.  
The observed bistability, and hysteretic behavior of the SQN embedded in a low-dissipative resonator may be exploited for realizing quantum threshold detectors or bistable memory elements in circuit quantum electrodynamics architectures.


\section*{Materials and methods}
\subsection*{A. Device Fabrication}
The fabrication of T- and R- resonators was performed by patterning by Reactive Ion Etching (RIE) a 200 nm thick Nb film deposited on a silicon substrate. The flux qubits were realized in Al by two-angle shadow evaporation technique, forming three Josephson junctions loops having lateral dimensions equal to 6.0x4.5 $\mu$m$^2$. Each loop incorporates two identical junctions (0.2x0.87 $\mu$m$^2$), while the third one is scaled of a factor 0.8. Each qubit is shunted by a coplanar capacitor. The critical current density of the fabricated network qubit is equal to 80 A/cm$^2$. Additional details on the fabrication process and on the flux qubit parameters are reported in~\cite{qn,App}.

\subsection*{B. Experimental setup}
To perform the scattering parameters measurements, the SQN network was mounted on a sample holder provided with six SMA connector for both RF and DC input and output signals. A superconducting NbTi filament was wound around the holder to provide the static magnetic field. The microwave signals were provided by a Rohde\&Schwartz SMA100B (applied to Port 3 in the scheme shown in Fig.~\ref{fig:2p}), while the transmission spectra of both T- and R- resonators were acquired by an Agilent E5071C VNA. The signal generator and VNA outputs were direc2pted to two RF lines into a Leiden cryostat. In particular, thermal noise coming from the hottest plates and from the environment was attenuated by two 20 dB attenuators located at the 4 K stage and at the 10 mK stage. Including also the losses introduced by the coaxial cables, the total attenuation of the system is about 60 dB in the operation frequency range. 

On the output side, cryogenic circulators (operating from 4 to 12 GHz) directed the signal from either the T- or R-resonator to a cryogenic switch, which routed it to the amplification stage. The signal was first amplified by a low-noise HEMT amplifier at 4 K, followed by a room-temperature FET amplifier, yielding a total gain of 60dB. Finally, a splitter distributes the output power to the VNA for the S$_{21}$ measurements and to the Signal Hound SM200B spectrum analyzer for power-spectrum measurements, as shown in Fig.~\ref{fig:3}. Additional details on the experimental setup scheme are reported in~\cite{qn}.

\subsection*{C. Derivation of cavity field equation}

\subsubsection*{C.1. Basic equations}
Our approach relies on the formalism for a cavity derived in the supplementary materials in \cite{qed}, with the main difference that, a result of mutual coupling, we consider the two  cavity R+T as one effective resonator of frequency $\omega_c$.
At port 3, the input electromagnetic field in a waveguide in the $z$ direction with the  coordinate $r_z$ can be represented by the field for  forward $+$ and backward $-$ microwave field
\begin{eqnarray}\label{Apm}
A_\pm  (r_z,t)&=&\frac{ V(r_z,t) \pm Z  I(r_z,t)}{2\sqrt{Z}}.
\end{eqnarray}
These are the  classical voltage and current usually associated to a coplanar  waveguide. $Z$ is the impedance of the line (and usually fixed to $60Hz$ for quantum circuit).
The input field is a free propagating field while the scattered one results from the interaction with the resonator and the qubits.  Around the pump frequency $\omega_0$ viewed also as a carrier frequency, we rewrite quite generally  the field enveloppes caused by an amplitude modulation (or frequency modulation on convenience) at the cavity end $r_z =0$ on port 3 :
\begin{eqnarray}\label{x}
A_- (0,t)= \sqrt{2 \hbar \omega_c} {\rm Re}(x_{out}(t)e^{-i\omega_0 t}) \\
A_+ (0,t)= \sqrt{2 \hbar \omega_c}{\rm Re}(x_{in}(t)e^{-i\omega_0 t} )\, .
\end{eqnarray}
The transmon  is  placed  inside the  cavity end for a maximum coupling to the field.
Adding the dephasing $\Gamma_{\phi,j}$ due to Josephson junction and the decay rate $\Gamma_j$ to a transverse channel (due to other cavity modes), we obtain the master equation for the matrix density $\hat \rho(t)$ \cite{qed}:
\begin{eqnarray}\label{mast2}
&&\partial_t \hat \rho(t) +i[\hat H_S(t),\hat \rho(t)]
\nonumber \\&=&
\frac{\gamma_c}{2} \left[2\hat a  \hat \rho(t) \hat a^\dagger -\{\hat a^\dagger \hat a, \hat \rho(t)\}\right]- \sum_{j=1}^N(\frac{\Gamma_j}{2}+\Gamma_{\phi,j})\{\hat \sigma^+_j \hat \sigma^-_j,\hat \rho(t)\}
-\Gamma_j\hat \sigma_j^- \hat \rho(t) \hat \sigma_j^+
  \,  .
  \nonumber \\
\end{eqnarray}
where $[.,.]$ and $\{., .\}$ are the commutator and the anticommutator respectively.
The associated system Hamiltonian written in the rotating wave approximation (RWA) is:
\begin{eqnarray}\label{Hs}
\hat H_S(t)&=&
-\omega_{0c}\hat a^\dagger \hat a
+i \sqrt{\frac{\omega_c}{2|Q_c|}}(x_{in}^*(t)\hat a - x_{in}(t)\hat a^\dagger ) 
\nonumber \\
&-&\sum_{j=1}^N
\left[\frac{\Delta_j}{2}\hat \sigma^z_j
+g_j(\hat a^\dagger \hat \sigma^-_j+\hat a \hat \sigma^+_j) \right]\, ,
\end{eqnarray}
where we define the Pauli matrices, describing the jth qubit, $\hat \sigma^{x,y,z}_j$, and
the Bose operators $\hat a^\dagger,\hat a$.
The lifetime for
the photon within the T+ R-resonator $T_{ph}=1/\gamma_c$ should include also the internal losses $\omega_c/Q_i$ in the relaxation rate $\gamma_c=(1/|Q_c| +1/Q_i)\omega_c$ of photon leakage in (\ref{mast2}) and differs from \cite{qed} where  $Q_i \rightarrow \infty$.
It is assumed to be much shorter
to the relaxation times for decay $T_{1,j}=1/\Gamma_j$ and for pure dephasing $T_{2,j}=1/\Gamma_{\phi,j}$ since, due to the Purcell effect,  the qubits interact mainly with the cavity field mode.
Their relaxation rates will be  therefore  neglected in the following compared to the leakage rate $\gamma_c$.

The master equation (\ref{mast2})  describes
only the dynamics of qubit due to an external field.
However, we need to specify another relation with the scattered field generated from the qubit evolution. For this purpose, we use  the Fourier transform in  time of the field as
\cite{qed}:
\begin{eqnarray}\label{ft}
{A}_\pm(r_z,t)=
\int_{-\infty}^\infty
\frac{d\omega}{2\pi}e^{-i(\omega+i\eta) t}
{A}_{\omega,\pm}(r_z)
\end{eqnarray}
where we have introduced the positive and infinitesimal parameter $\eta \rightarrow 0$ to ensure an adiabatic switching essential to study the retarded response to any input wave.

Restricting the frequency domain close to the cavity mode, the reflection is (up to a prefactor neglected here):
\begin{eqnarray}\label{S33a}
S_{33} =
\frac{ A_{\omega_0,-}(0)}{ A_{\omega_0,+}(0)}
=1+\sum_{\pm}
\frac{e^{i\phi} Q_l \gamma_c/2}{\omega_0 -(\pm \omega_c - i\gamma_c/2)}
\left[\frac{i}{|Q_c|} \pm
\sqrt{\frac{\hbar}{|Q_c|} }\sum_{j=1}^N  \frac{g_j\langle \hat \sigma_{\omega_0,j}^\mp \rangle}{A_{\omega_0,+}(0)} \right]  \,  .
\end{eqnarray}
We include phenomenologically the internal loss with finite internal quality factor and
the  mismatch phase factor $\phi$ \cite{probst}, in contrast to  \cite{qed} where we use $\phi=\pi$.

\subsubsection*{C.2. Semi-classical solution}

Using the Glauber-Sudershan transformation $\hat D(t)$, we rewrite the cavity field as:
\begin{eqnarray}\label{beta0}
\hat a \rightarrow  \hat D^\dagger(t) \hat a \hat D(t)=\hat a+\psi(t)
\end{eqnarray}
where $\psi(t)$ is the coherent field inside the resonator while the  new $\hat a$ defines now the quantum fluctuations operator.
Similarly we can redefine the qubit operators adequately in the dressed state representation. Using the polar decomposition $\psi(t)=e^{i\theta(t)} \sqrt{n_{ph}(t)}$
we define
\begin{eqnarray}
\hat \sigma^x_j= e^{-i\theta(t)} \hat \sigma^-_j+ e^{i\theta(t)}\hat \sigma^+_j
\\
\hat \sigma^y_j= i(e^{-i\theta(t)} \hat \sigma^-_j- e^{i\theta(t)}\hat \sigma^+_j)
\end{eqnarray}
After an additional rotation about the y-axis in the qubit Bloch sphere through
\begin{eqnarray}
\hat \sigma'^z_j&=&\cos \theta_j(t) \hat \sigma^z_j +\sin \theta_j(t) \hat \sigma^x_j
\\
\hat \sigma'^x_j&=&-\sin  \theta_j(t) \hat \sigma^z_j +\cos \theta_j(t) \hat \sigma^x_j
\\
\cos \theta_j(t) &=& \frac{|\Delta_j|}{\sqrt{\epsilon^2_j(t)+\Delta^2_j}}
\quad \quad \sin \theta_j(t) = \frac{\Delta_j \epsilon_j(t)/|\Delta_j|}{\sqrt{\epsilon^2_j(t)+\Delta^2_j}}
\nonumber \\
\end{eqnarray}
where $\epsilon_j(t)=2 g_j |\psi(t)|$ is the Rabi frequency of qubits $j$.
Using these successive time-dependent unitary transformation, the Hamiltonian has the new form (omitting the time dependence notation):
\begin{eqnarray}\label{int}
&&\hat H'_S =
\left[\left(i\frac{d}{dt} +\omega_{0c} -\sum_{j=1}^N \frac{g_j^2/\Delta_j}{\sqrt{1+\epsilon^2_j/\Delta^2_j}}\right)
\psi\hat a^\dagger +i \sqrt{\frac{\omega_c}{2|Q_c|}}x_{in}\hat a^\dagger+c.c.\right]
\nonumber \\
&-&
\sum_{j=1}^N\bigg[\frac{\sqrt{1+\epsilon^2_j/\Delta_j^2}}{2}\Delta_j(\hat \sigma'^z_j +1)
+\frac{g_j(\epsilon_j (\hat \sigma'^z_j +1)+\Delta_j \hat \sigma'^x_j)}{2\Delta_j\sqrt{1+\epsilon^2_j/\Delta^2_j}} (e^{-i\theta}\hat a + e^{i\theta}  \hat a^\dagger)  
\nonumber \\
&+&\frac{ig_j}{2}\hat \sigma^y_j(e^{-i\theta}\hat a - e^{i\theta}  \hat a^\dagger) \bigg]
\nonumber \\
\end{eqnarray}
where we define the Landau-Ginzburg-like function
\begin{eqnarray}
{\cal{F}}(\psi,\psi^*)&=& i(\frac{d\psi^*}{dt}\psi-\frac{d\psi}{dt}\psi^*)-
\omega_{0c}|\psi|^2+ \frac{1}{2} \sum_{j=1}^N \Delta_j \sqrt{1+4g^2_j|\psi|^2/\Delta^2_j}
\nonumber \\
&+&
i \sqrt{\frac{\omega_c}{2|Q_c|}} (x_{in}^*\psi - x_{in}\psi^* )  \, ,
\end{eqnarray}
In absence of cavity leakage, the latter corresponds to a Lagrangian allowing alternatively to recover the  dynamic equation (\ref{DynamicsEq}). It is an analog of the Gross-Pitaevski equation but with a different form for the non-linearity.
To obtain (\ref{int}), we have neglected the derivative terms $d\theta /dt$ and
$d\theta_j/dt$ resulting from the time dependent unitary transformation as they are assumed small ($\sim 2\pi$ MHz)  in comparison to the detuning $\Delta_j$ ($\sim 2\pi 100$ MHz).
These transformations redefines  the density matrix $\hat \rho$  into $\hat \rho'$ as well as the  master equation which becomes:
\begin{eqnarray}\label{mast2p}
&&\partial_t \hat \rho' +i[\hat H'_S,\hat \rho']
=
\frac{\gamma_c}{2} \left[2(\hat a+ \psi)  \hat \rho'  (\hat a^\dagger+\psi^*) -\{(\hat a^\dagger+\psi^*)(\hat a+ \psi), \hat \rho'\}\right]
  \,  .
\end{eqnarray}
We cancel out all the linear terms in the creation annihilation operator in Eq.(\ref{mast2p}) by imposing the condition of Eq.(\ref{DynamicsEq}).
The remaining contributions are negligible
in the semi-classical approximation when the coherent field is important ($|\psi|^2 \gg 1$) in comparison to quantum fluctuations. Therefore we omit the quantum part of the cavity photon field in the interaction with the qubits and  the master equation  reduces more simply to:
\begin{eqnarray}\label{mast3}
&&\partial_t \hat \rho' -i[\omega_{0c} \hat a^\dagger \hat a +
\sum_{j=1}^N\frac{\sqrt{1+\epsilon^2_j/\Delta_j^2}}{2}\Delta_j(\hat \sigma'^z_j +1),\hat \rho']
=
\frac{\gamma_c}{2} \left[2\hat a  \hat \rho' \hat a^\dagger -\{\hat a^\dagger \hat a, \hat \rho'\} \right]
  \,  .
  \nonumber \\
\end{eqnarray}
It has the trivial  ground state solution where $\langle  \hat a^\dagger \hat a \rangle=0 $ and $\langle \hat \sigma'^z_j \rangle = -1$. Returning to the initial representation, we deduce:
\begin{eqnarray}\label{sig}
 \langle \hat \sigma^-_j \rangle =-\frac{2 g_j \psi/\Delta_j }{\sqrt{1+ 4 g^2_j n_{ph}/\Delta^2_j}}
\end{eqnarray}
Therefore the dynamics is mainly  governed by the  Eq.(\ref{DynamicsEq}) in the semi-classical case.
In the case of the steady state solution where $x_{in}=x_0$ is time-independent, we find
\begin{eqnarray}\label{b0}
\psi= \frac{-ix_{0}\sqrt{\omega_c/2|Q_c|}}{\omega_s +i\gamma_c/2} 
\end{eqnarray}
For proving (\ref{eq:luca}), we note the relation $A_{\omega,+}(0)
=\sqrt{2\hbar\omega_c}x_{in}$  that we combine algebraically together with
Eqs.(\ref{sig},\ref{b0},\ref{S33a}).

\subsection*{D. $S_{33}$ parameter estimation}

In this section, we describe the methods used to estimate the loaded quality factor $Q_l$ and coupling quality factor $Q_c$ from  of the $S_{33}$  resonator model by using the fitting procedure of the  experimental data. The process involves several steps, including data preprocessing, interpolation, model definition, fitting, and evaluation of results.

\begin{figure}[H]
    \centering
    \includegraphics[width=0.8\textwidth]{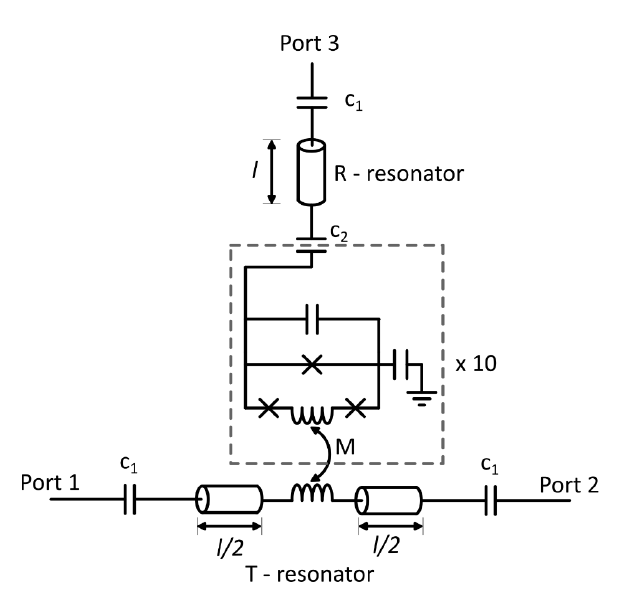}
    \caption{Equivalent circuit of the SQN setup.}
    \label{fig:resonator}
\end{figure}

The SQN setup has been modeled as shown in Fig.~\ref{fig:resonator}. It consists of two finite-length transmission lines acting as microwave resonators. The readout resonator, spanning a total length \textit{l}, is defined between ports 1 and 2, while the sensing resonator, also of length \textit{l}, is connected to port 3  . Both resonators are capacitively terminated at their ends and support standing-wave modes at frequencies corresponding to integer multiples of \textit{(l/2)}.
The T- geometry arrangement of the two resonators ensures weak mutual coupling in the absence of the flux qubit network~\cite{qn}  .

\begin{figure}[!htbp]
    \centering
    \includegraphics[width=1\linewidth]{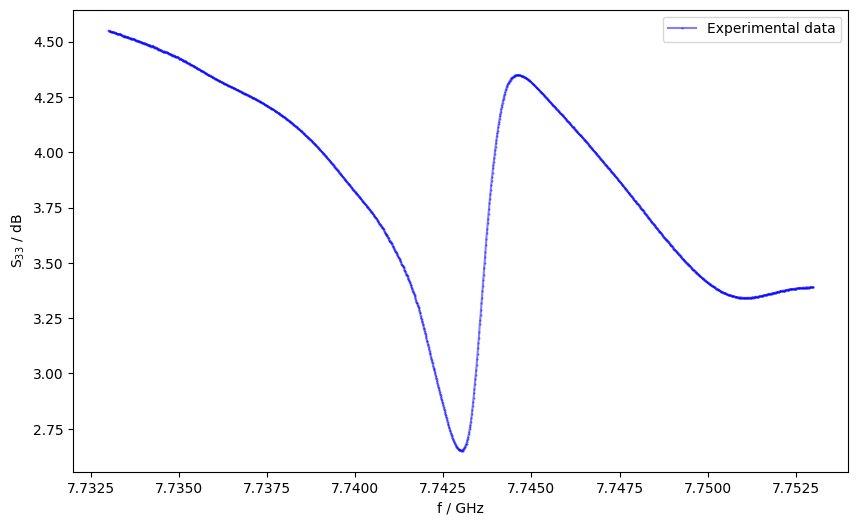}
    \caption{Plot of the experimental data showing the $S_{33}$ as a function of frequency.}
    \label{fig:exp_data}
\end{figure}

\begin{figure}[!htbp]
    \centering
    \includegraphics[width=1\linewidth]{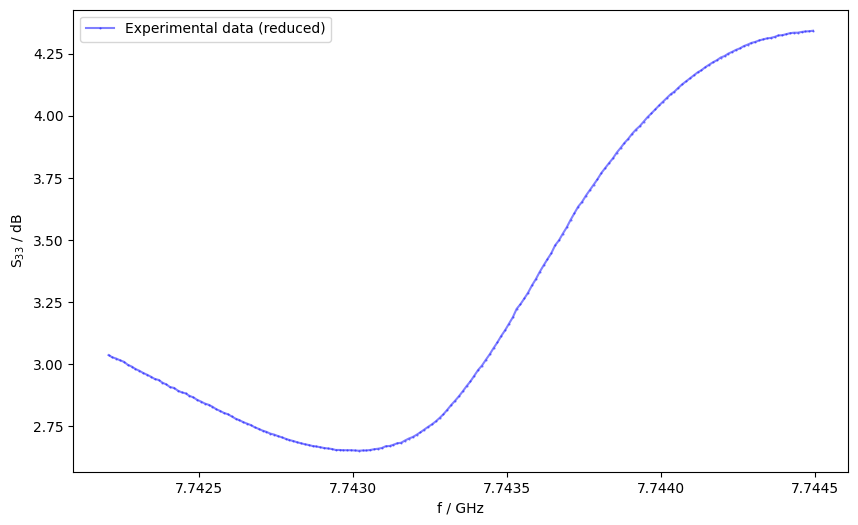}
    \caption{Plot of the experimental data within the frequency window $[7.7422, 7.7445]$ GHz.}
    \label{fig:red_exp_data}
\end{figure}

In the first step, we preprocessed the data by loading the experimental measurements of $S_{33}$ as a function of frequency from a raw file (Fig.~\ref{fig:exp_data}). The data was filtered to include only measurements within a specified frequency window, focusing on the interval of interest around the resonance dip. In this case, the frequency window used was $[7.7422, 7.7445]$ GHz (Fig.\ref{fig:red_exp_data}). 

Next, we interpolated the reduced data using a cubic spline. This interpolation created a smooth curve that better represents the experimental data, providing a continuous representation essential for an accurate fitting.

We then use equation \eqref{eq:luca} 
related to $S_{33}$ Notch Resonator model to describe the behavior of $S_{33}$ as a function of the resonance frequency ($f_r$), quality factors ($Q_c$ and $Q_l$), amplitude ($a$), and phase ($\phi$) as follows:

\begin{equation}
S_{33} = a \cdot \left( 1 - \frac{ \dfrac{Q_\ell}{|Q_c|} e^{i\phi} }{ 1 + 2i Q_\ell \left( \dfrac{f}{f_r} - 1 \right) }  \right)
\label{eq:luca}
\end{equation}
Note in this expression the opposite imaginary sign convention of $i$ used in microwave engineering  that differs from the one used in quantum mechanics (see \cite{pozar} where the imaginary unit $j$ is opposite in our notation $i=-j$).


The loaded quality factor $Q_l$ can also be expressed in terms of the internal quality factor $Q_i$ as follows:

\begin{equation}
\frac{1}{Q_l} = \left| \frac{1}{Q_c} \right| + \frac{1}{Q_i}
\label{eq:ql_def}
\end{equation}

In energy terms, $Q_l$ represents the ratio between the energy stored in the resonator and the energy lost per period $(\tau = 1/ f_r)$ due to all the sources of dissipation (both those internal to the resonator mainly fluorescence (see paper in preparation) and those caused by coupling with the external environment). Note that for the second tone of weak intensity, the internal loss is negligible and therefore $Q_l \simeq Q_c$.

\begin{figure}[H]
    \centering
    \includegraphics[width=1\linewidth]{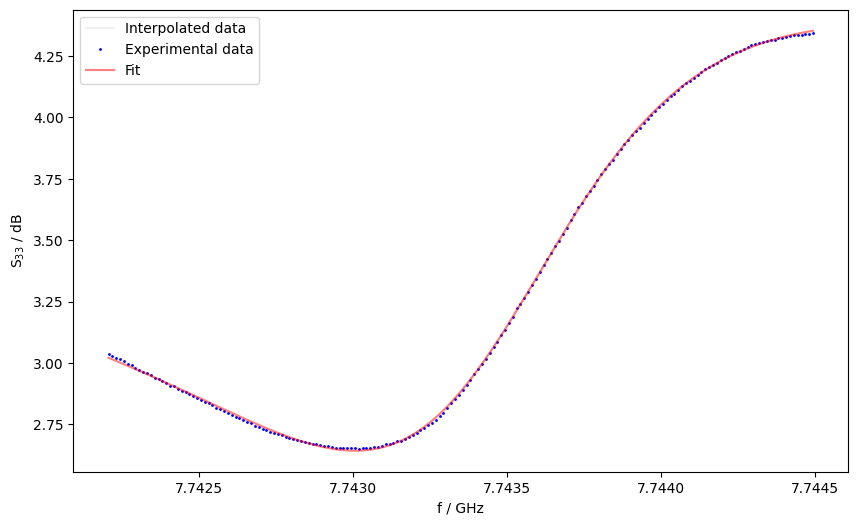}
    \caption{Plot of the experimental data, interpolation, and the obtained fit.}
    \label{fig:fit}
\end{figure}

We used \textit{Powell's Dog Leg} fitting method (\cite{num}) to fit the model to the interpolated data (Fig.~\ref{fig:fit}). This method optimizes the model parameters to minimize the discrepancy between the experimental data and the theoretical model. The fitting was performed using the \texttt{scipy} (\cite{scipy}) library, which identified the optimal parameters that best fit the data.

The fitting results included the optimal model parameters and their uncertainties. In addition, the Root Mean Square Error (RMSE) was calculated to estimate the quality of the fit. The optimal parameters obtained are:

\begin{lstlisting}[mathescape]
    $Q_c = 24530.93 \pm 0.38$
    $Q_l = 4682.71 \pm 0.05$
    $a = 1.574 \pm 0.00002$
    $f_r = 7.743538 \pm 0.000000$
    $\phi = -1.054 \pm 0.00002$
\end{lstlisting}

The RMSE for the interpolated data was $RMSE_{int} = 0.005348$, while for the experimental data it was $RMSE_{exp} = 0.005453$.

\begin{figure}
    \centering
    \includegraphics[width=1\linewidth]{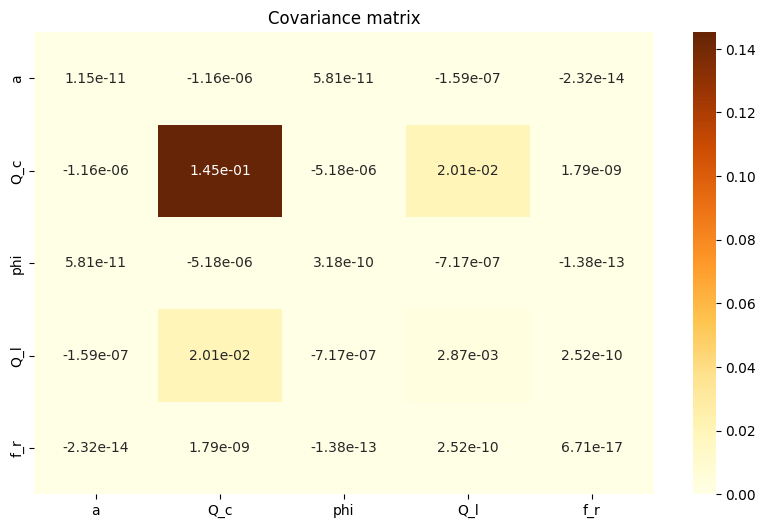}
    \caption{Covariance matrix of the fitted parameters.}
    \label{fig:cov_matrix}
\end{figure}

Finally, the covariance matrix obtained from the fitting process is shown in Figure~\ref{fig:cov_matrix}.

\end{document}